\documentclass[conference]{IEEEtran}

\IEEEoverridecommandlockouts

\usepackage{cite}
\usepackage{amsmath}
\usepackage{cuted}
\usepackage{amsopn}

\usepackage{float}

\usepackage{dblfloatfix}

\usepackage{setspace}

\usepackage{amssymb}

\usepackage{amsfonts}

\usepackage{graphicx}
\usepackage{diagbox} 

\usepackage{booktabs}

\usepackage{caption3} 
\DeclareCaptionOption{parskip}[]{} 
\usepackage[small]{caption}

\usepackage{geometry}
\usepackage{setspace,graphicx,multirow}
\usepackage{subcaption}
\usepackage{array}
\usepackage{algorithm}
\usepackage[]{algpseudocode}
\makeatletter
\def\BState{\State\hskip-\ALG@thistlm}
\makeatother
\usepackage{algcompatible}
\usepackage{url}

\usepackage{multirow}
\usepackage{hhline}
\usepackage{xspace}

\usepackage{color}

\ifodd 1
\newcommand{\com}[1]{\textbf{\color{blue} (COMMENT: #1)}} 
\else

\newcommand{\com}[1]{}
\fi

\usepackage{enumitem}
\setenumerate[1]{itemsep=0pt,partopsep=0pt,parsep=\parskip,topsep=0pt}
\setitemize[1]{itemsep=0pt,partopsep=0pt,parsep=\parskip,topsep=0pt}
\setdescription{itemsep=0pt,partopsep=0pt,parsep=\parskip,topsep=0pt}

\begin{document}
\bibliographystyle{IEEEtran}
\bstctlcite{IEEEexample:BSTcontrol}
 
\title{Neural Channel Knowledge Map Assisted Scheduling Optimization of Active IRSs in Multi-User Systems}

\author{Xintong~Chen,
        Zhenyu~Jiang,
        Jiangbin~Lyu,~\IEEEmembership{Member,~IEEE},
        and~Liqun~Fu,~\IEEEmembership{Senior Member,~IEEE}
\thanks{The authors are with the School of Informatics, Xiamen University (XMU), China. \textit{Corresponding author: Jiangbin Lyu} (email: ljb@xmu.edu.cn).
}
}

\maketitle


\begin{abstract}
Intelligent Reflecting Surfaces (IRSs) have potential for significant performance gains in next-generation wireless networks but face key challenges, notably severe double-pathloss and complex multi-user scheduling due to hardware constraints. Active IRSs partially address pathloss but still require efficient scheduling in cell-level multi-IRS multi-user systems, whereby the overhead/delay of channel state acquisition and the scheduling complexity both rise dramatically as the user density and channel dimensions increase.
Motivated by these challenges, this paper proposes a novel scheduling framework based on \textit{neural} Channel Knowledge Map (CKM), designing Transformer-based deep neural networks (DNNs) to predict ergodic spectral efficiency (SE) from historical channel/throughput measurements tagged with user positions. Specifically, two cascaded networks, LPS-Net and SE-Net, are designed to predict link power statistics (LPS) and ergodic SE accurately. We further propose a low-complexity Stable Matching-Iterative Balancing (SM-IB) scheduling algorithm. Numerical evaluations verify that the proposed neural CKM significantly enhances prediction accuracy and computational efficiency, while the SM-IB algorithm effectively achieves near-optimal max-min throughput with greatly reduced complexity.
\end{abstract}


\IEEEpeerreviewmaketitle

\section{Introduction}

Intelligent reflecting surface (IRS), a.k.a. reconfigurable intelligent surface (RIS), holds promise in next generation wireless networks for significantly enhanced performance, which attracts extensive research on various aspects such as hardware/beamforming design, channel modelling/estimation, and network analysis/optimization (see, e.g., \cite{RecentAdvances} and the references therein).
Nevertheless, two key issues pose limits to the achievable performance gains in IRS-aided systems, i.e., 1) the double-pathloss effect of IRS-reflected channel constrains its service range for distant user equipments (UEs), and 2) the typical time-selective nature\cite{ZhengBeiXiongNOMAorOMA} of IRS hardware requires dedicated scheduling when serving multiple UEs.

To compensate for the double-pathloss, large IRS with massive elements could be employed, which yet increases the hardware cost and also computational complexity for phase optimization. 
An alternate solution is to exploit the recently proposed active IRS (AIRS)\cite{zhang2021active}, which helps alleviate the power loss in IRS-reflected channel by incorporating power amplifiers to amplify/control the reflection amplitude in addition to phase.
Though at the cost of higher hardware complexity and energy consumption, the advantages of AIRS over passive IRS have been advocated by recent studies\cite{ActiveIRSOpenIssue, PanFormula}, especially for cell-level coverage enhancement\cite{Dongsheng}.
Despite the link-level coverage extension, the full potential of AIRS in multi-user systems depends not only
on the AIRS beamforming, but also on the scheduling optimization of UE-to-AIRS association and time/frequency resource allocation, especially considering the time sharing constraint per AIRS when serving multiple UEs.

Scheduling optimization in multi-IRS aided multi-user systems need dedicated design \cite{OFDMA-schedule-4, OFDMA-schedule-5, NOMA-schedule-1}.
In compliance with 5G standards, the works in \cite{OFDMA-schedule-4,  OFDMA-schedule-5} employ orthogonal frequency division multiple access (OFDMA) to support multi-user transmissions by joint subchannel allocation/power control/phase optimization.
In addition, resource allocation in IRS-aided non-orthogonal multiple access (NOMA) schemes have also been investigated (see, e.g., \cite{ZhengBeiXiongNOMAorOMA,NOMA-schedule-1} and the references therein).
The aforementioned schemes typically assume the availability of instantaneous channel state information (CSI) of all UEs involved in the scheduling.
However, as UE density and channel dimensions increase, the overhead and delay of obtaining full CSI rise dramatically, rendering the scheduling decisions obsolete/ineffective, not to mention the complexity of scheduling which is typically discrete and NP-hard in nature.

In this paper, we tackle the above challenges by leveraging channel knowledge map (CKM)\cite{CKMTutorial} for statistical channel inference/performance prediction.
By exploiting various positioning/sensing technologies with ever increasing accuracy,
CKM aims to map any possible UE location to its associated channel knowledge or performance metrics.
Essentially, CKM accumulates historical channel/performance measurements into spatially correlated databases, which has the unique potential for (statistical) channel/performance inference across different time/frequency/space/devices and thus circumvents real-time full CSI acquisition.
Note that the validity of CKM for IRS-aided systems has been consolidated in application scenarios such as beam selection\cite{CKM_IRSbeamSelection} and joint active/passive beamforming\cite{CKM_ActivePassive},
whereas multi-IRS scheduling in cell-level multi-user systems is yet to be considered.
This poses further challenges for CKM construction/inference due to the high-dimensional configuration space for multiple IRSs,
for which the conventional table-based CKM\cite{CKMTutorial} suffers from excessive storage and difficulty of multi-conditional inference.

Motivated by the above, we consider a single-cell multi-user system aided by multiple AIRSs and aim to maximize the minimum ergodic throughput among all UEs, by jointly optimizing the UE-to-AIRS association and time/frequency resource allocation subject to the AIRS time sharing constraints.
Note that we choose ergodic throughput instead of instantaneous rate as the objective, due to the above mentioned overhead and delay of obtaining instantaneous CSI of all UEs which might obsolete the scheduling.
However, due to the lack of analytical formulas for ergodic spectrum efficiency (SE) or throughput, we need practical ways to evaluate SE under given scheduling.
To this end, we propose a \textit{neural CKM} method based on deep neural networks (DNNs) for efficient storage and prediction of link power statistics (LPS) and further the ergodic SE in AIRS-aided systems.
Specifically, similar to \cite{PanFormula}, assume that the IRS-related channel power statistics could be obtained through statistical channel measurements, while the UE's ergodic SE under given scheduling could be estimated by statistical average of throughput measurements.
The neural CKM consists of two cascaded DNNs named LPS-Net and SE-Net, each responsible for 1) predicting the per-link power statistics based on given UE positions and system configurations; and 2) composing the link power statistics to infer the ergodic SE under given UE-to-AIRS association, respectively.
To capture the highly nonlinear function mappings associated with these two sub-tasks, we choose Transformer\cite{viT} as the core building blocks for flexible input encoding and output mapping, which is compared favorably against baselines based on multi-layer perceptron (MLP) and long short-term memory (LSTM).
Finally, based on neural CKM for ergodic SE prediction, we further propose a Stable Matching-Iterative Balancing (SM-IB) algorithm for efficient AIRS and time/frequency scheduling, whereby the achieved max-min ergodic throughput is close to the upper bound obtained by the Gurobi solver, yet with significantly reduced running time.

\section{System Model}\label{Sec:SystemModel}

\begin{figure} [t]
	\centering
	\includegraphics[width=0.55\linewidth,  trim=0.0 20.0 0.0 15.0, clip]{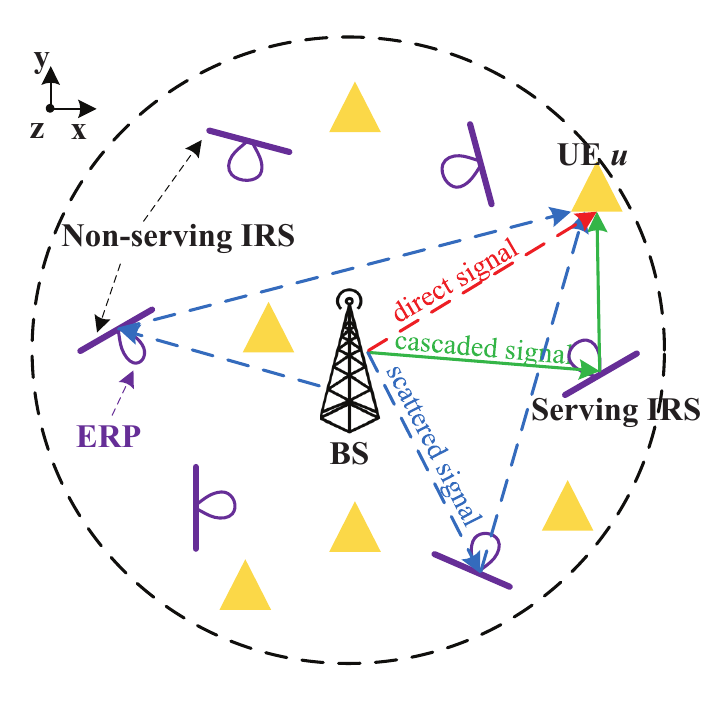}
	\caption{Multi-AIRS assisted multi-user systems.\vspace{-2ex}}
    \label{Fig:system_model}
\end{figure}

We consider a multiuser downlink communication network assisted by multiple AIRSs in a single cell, where the BS is located at the origin with a height of $H_\text{BS}$ m, as in Fig. \ref{Fig:system_model}.
Within the cell, $I$ AIRSs are deployed, denoted by the set $\mathcal{I} \triangleq \left\{1,...,I \right\}$, each comprising $W\triangleq W_\text{Y}\times W_\text{Z}$ reflecting elements as a set $\mathcal{W}$. 
Each IRS panel $i\in\mathcal{I}$ is placed at the location $\left(x_i, y_i, H_i\right)$ with counterclockwise rotation angles $\omega_{\text{x}, i}$, $\omega_{\text{y}, i}$ and $\omega_{\text{z}, i}$ about the $X$, $Y$, and $Z$ axes, respectively, based from the default Y-O-Z plane.
Denote the set of randomly distributed UEs as $\mathcal{U} \triangleq \left\{1,..., U\right\}$, where the position of UE $u \in \mathcal{U}$ is given by $\left(x_u, y_u, H_u\right)$.
For simplicity, the BS and UEs are assumed to be equipped with a single isotropic antenna.\footnote{We focus on the modeling of AIRSs in this work, while the proposed method can be extended to the scenarios with multi-antenna BS and UEs.}
On the other hand, the AIRS is assumed to exhibit a certain element radiation pattern (ERP)\cite{TransShiJinPathlossmodeling}, denoted by $G_{i}\left(\Omega_{x,i}\right), x \in \left\{\text{t},\text{r}\right\}$, which represents the power pattern from/to the azimuth and elevation angles $\Omega_{x,i} \triangleq \left(\theta_{x,i}, \varphi_{x,i}\right)$, i.e.,
\begin{small}
\begin{equation}
G_{i}\left( \Omega \right) \triangleq 
\left\{
\begin{array}{cc}
G\left(\sin\theta \cos\varphi\right)^{q_i}, & \theta \in [0,\pi], \varphi \in [-\frac{\pi}{2},\frac{\pi}{2}], \\
0, & {\rm otherwise},
\end{array}
\right.
\end{equation}%
\end{small}%
where $G$ denotes the maximum element power gain, and $q_i$ controls the shape of the ERP.

Orthogonal frequency division multiple access (OFDMA) is adopted to mitigate multiuser interference, where the total bandwidth $B$ is equally divided into $S$ resource blocks (RBs), comprising a set $\mathcal{S}$, and the time frame $T$ is divided into $Q$ slots, comprising a set $\mathcal{Q}$.
Each UE $u$ is assigned a set of orthogonal RBs $\mathcal{S}_{u,q}\subseteq \mathcal{S}$ in a given slot $q\in\mathcal{Q}$. 
To maximize the beamforming gain of each IRS to each of its served UEs, we assume that they are assigned in orthogonal-time RBs, i.e., at each time slot, each IRS serves at most one UE\cite{ZhengBeiXiongNOMAorOMA}.

\vspace{-1ex}

\subsection{Channel Model}

For RB $s$, the channels from the BS to UE $u$, from the BS to AIRS $i$, and from AIRS $i$ to UE $u$ are denoted as $h_{\text{b},u,s} \in \mathbb{C}$, $\mathbf{h}_{\text{b},i,s} \in \mathbb{C}^{W \times 1}$, and $\mathbf{h}_{i,u,s} \in \mathbb{C}^{1 \times W}$, respectively.
To be aligned with the standard 3GPP channel evaluation assumptions, we adopt the QuaDRiGa toolbox to generate the underlying channels.
Specifically, since the deployment of AIRSs typically guarantees a line-of-sight (LoS) link between the BS and each AIRS, $\mathbf{h}_{\text{b},i,s}$ is modeled according to the "3GPP-38.901-UMa-LoS" scenario.
Due to the uncertainty in UE positions, a non-LoS (NLoS) link between the BS/AIRS and each UE is more common. Therefore, the "3GPP-38.901-UMa-NLoS" scenario is adopted for $h_{\text{b},u,s}$ and $\mathbf{h}_{i,u,s}$.

As a result, the BS-AIRS $i$-UE $u$ cascaded channel at RB $s$ is given by
\begin{equation}\label{cascadedChannel}
    h_{\text{b},i,u,s} \triangleq \mathbf{h}_{i,u,s} \mathbf{F}_i \mathbf{\Phi}_i \mathbf{h}_{\text{b},i,s},
\end{equation}
where $\mathbf{F}_{i} \triangleq \text{diag}\left(F_{i,1}, ..., F_{i,W}\right) \in \mathbb{R}^{W\times W}$ is the amplification factor, and $\mathbf{\Phi}_{i} \triangleq \text{diag}\left(\phi_{i,1}, ..., \phi_{i,W}\right) \in \mathbb{C}^{W\times W}$ is the unit-modulus phase coefficient of AIRS $i$.
Due to active reflection, dynamic noise $\mathbf{v}_i\in\mathbb{C}^{W\times 1}$ with per-element power spectrum density (PSD) $N_\text{v}$ is generated during amplification\cite{zhang2021active}.
Thus, the received signal $r_{u,s}$ at UE $u$ and RB $s$ is given by
\begin{equation}
    r_{u,s} \triangleq \left( h_{\text{b},u,s} + \textstyle\sum\nolimits_{i\in\mathcal{I}}h_{\text{b},i,u,s} \right) g_{u,s} + \textstyle\sum\nolimits_{i\in\mathcal{I}} \mathbf{h}_{i,u,s} \mathbf{F}_i \mathbf{\Phi}_i \mathbf{v}_i + n_0,
\end{equation}
where $g_{u,s}$ is the transmit signal with transmit power $p_s$, and $n_0$ is the receiver thermal noise with PSD $N_\text{0}$.


For simplicity, assume that each UE $u$ is served by at most one AIRS in each time slot.
The signal reflected from the serving AIRS to the UE is termed as \textit{cascaded signal}, while signals reflected from non-serving AIRSs are referred to as \textit{scattered signals}, as depicted in Fig. 1.
In cases without serving AIRS, all AIRSs simply reflect randomly scattered signals.
As a result, 
the received signal-to-noise ratio (SNR) of UE $u$ at RB $s$ served by AIRS $j$ is given by
\begin{equation}\label{SNR}
    \gamma_{u,s,j}\triangleq \frac{
            p_s \left| h_{\text{b},u,s} + h_{\text{b},j,u,s} + \sum_{i=1, i\neq j}^{I} h_{\text{b},i,u,s} \right|^2
        }{
           \sum_{i=1}^{I} \left\| \mathbf{h}_{i,u,s} \mathbf{F}_{i} \mathbf{\Phi}_{i} \right\|^2 \sigma^2_\text{v} + \sigma^2_\text{0}
        },
\end{equation}
which is varying under random channel fading, with $\sigma^2_\text{v}=N_\text{v}B/S$ and $\sigma^2_\text{0}=N_\text{0}B/S$.
Therefore, the ergodic SE of UE $u$ in slot $q$ served by AIRS $j_{u,q}$ is given by 
\begin{equation}\label{SEFormula}
    \overline{\eta}_{u,q} \triangleq 
    \mathbb{E} \big\{
        \frac{1}{| \mathcal{S}_{u,q}|} \textstyle\sum\nolimits_{s\in\mathcal{S}_{u,q}}
        \log_2 \left( 1 + 
        \gamma_{u,s,j_{u,q}}\right)
    \big\}.
\end{equation}
The ergodic throughput of UE $u$ in bps/Hz in a time frame is then given by
\begin{equation}\label{Ru}
    R_u\triangleq \textstyle\sum\nolimits_{q\in\mathcal{Q}}| \mathcal{S}_{u,q}|\cdot\overline{\eta}_{u,q}=\textstyle\sum\nolimits_{q\in\mathcal{Q}} \rho_{u,q}S\cdot\overline{\eta}_{u,q},
\end{equation}
where $\rho_{u,q}\triangleq | \mathcal{S}_{u,q}|/S$ denotes the ratio of total RBs allocated to UE $u$ in slot $q$.

\subsection{AIRS Coefficient Optimization under Given Scheduling}\label{beamforming}
Given a certain UE-to-AIRS association, the amplitude and phase of each AIRS element can be optimized to enhance the UE performance.
For ease of practical implementation, assume that all elements share a common amplification factor, which is tuned to maximize the signal amplification, as in \cite{zhang2021active}.


Regarding phase optimization,
various IRS beamforming schemes exist with different level of optimality and complexity, e.g., \cite{CapacityZhangShuoWen} and \cite{JiangTaoMCCM}.
For practical implementation, we evaluate these algorithms in numerical studies, and adopt the method in \cite{JiangTaoMCCM} which achieves good performance with low complexity.
Specifically, the method in \cite{JiangTaoMCCM} aims to maximize the average cascaded signal power by employing the eigenvector corresponding to the largest singular value of the mean channel covariance matrix (MCCM).
The eigenvector is then projected to the unit manifold to extract the phase coefficients.

Finally, note that our proposed neural CKM method and further scheduling optimization are general and applicable under any given AIRS phase optimization schemes.

\section{Scheduling Optimization of Multiple AIRSs in Multi-User Systems} \label{schedule}
The overall performance in multi-AIRS aided multi-user systems depends not only on the AIRS coefficient optimization, but also on the scheduling optimization of UE-to-AIRS association and time/frequency RB allocation, especially considering the time sharing constraint per AIRS when serving multiple UEs.
Here, we first formulate an optimization problem to maximize the minimum ergodic throughput among all UEs.
Due to the lack of analytical formulas for ergodic SE or throughput, we further discuss practical ways to evaluate SE and motivate our proposed neural CKM method, which bypasses the online evaluation of instantaneous SE (and hence the associated overhead/delay) under different scheduling trials.

\subsection{Problem formulation}
Denote $a_{s,u,q}$ as the binary indicator for RB allocation, where $a_{s,u,q} = 1$ indicates that RB $s$ in slot $q$ is allocated to UE $u$.
Denote $\mu_{i,u,q}$ as the binary indicator for UE-to-AIRS association, where $\mu_{i,u,q} = 1$ indicates that AIRS $i$ serves UE $u$ in slot $q$.
Denote $p_{s,q}$ as the transmit power allocated to RB $s$ in slot $q$.
Denote $\phi_{i,q,w}$ as the phase coefficient of the $w$-th element of AIRS $i$ in slot $q$.
The overall optimization problem is formulated as follows:
\begin{align}
\mathrm{(P1)}:& \underset{
	\begin{subarray}{c}
	a_{s,u,q},\mu_{i,u,q}, \mu_{\text{d},u,q}, p_{s,q}, \phi_{i,q,w}\\
	u \in \mathcal{U}, i \in \mathcal{I}, w \in \mathcal{W}, q \in \mathcal{Q}, s \in \mathcal{S}
	\end{subarray}
}{\max} \quad \min_{u \in \mathcal{U}} R_u \notag\\
\text{s.t.}\quad
&\textstyle\sum\nolimits_{u\in\mathcal{U}} a_{s,u,q} \leq 1, \forall s \in \mathcal{S}, q \in \mathcal{Q}, \label{SubcarrierConstraintInP1} \\
&a_{s,u,q} \in \{ 0, 1\}, \forall u \in \mathcal{U}, s \in \mathcal{S}, q \in \mathcal{Q}, \label{SubcarrerVarInP1} \\
&\textstyle\sum\nolimits_{u\in\mathcal{U}} \mu_{i,u,q} \leq 1, \forall i \in \mathcal{I}, q \in \mathcal{Q}, \label{MatchConstraintInP1} \\
&\textstyle\sum\nolimits_{i\in\mathcal{I}} \mu_{i,u,q} \leq 1, \forall u \in \mathcal{U},  q \in \mathcal{Q}, \label{MatchConstraint2InP1} \\
&\mu_{i,u,q}\in \{ 0, 1\}, \forall u \in \mathcal{U}, i \in \mathcal{I}, q \in \mathcal{Q},  \label{MatchVarInP1} \\
&\textstyle\sum\nolimits_{s\in\mathcal{S}} p_{s,q} \leq P_\text{BS}, \forall q \in \mathcal{Q},  \label{PowerConstraintInP1} \\
&\left|\phi_{i,q,w}\right| = 1, \forall i \in \mathcal{I},  q \in \mathcal{Q}, w \in \mathcal{W}, \label{IRSConstraintInP1}
\end{align}
where \eqref{SubcarrierConstraintInP1} suggests that each RB is allocated to at most one UE; \eqref{MatchConstraintInP1} suggests that each AIRS serves at most one UE per slot; \eqref{MatchConstraint2InP1} suggests that each UE is served by at most one AIRS in each slot; 
\eqref{PowerConstraintInP1} represents the total transmit power constraint per slot;
and \eqref{IRSConstraintInP1} denotes the constant-modulus constraints on AIRS phase coefficients.
Note that the UE-to-AIRS association indicators $\mu_{i,u,q}$ affect the throughput $R_u$ in \eqref{Ru} by determining the serving AIRS $j_{u,q}$ and hence the ergodic SE in \eqref{SEFormula}.

Problem (P1) is a mixed-integer nonlinear program, which is difficult to solve especially for large number of UEs and AIRSs.
Moreover, due to random channel fading, there is no closed-form expression for the ergodic SE in \eqref{SEFormula} or throughput $R_u$ in \eqref{Ru}, 
rendering it difficult to evaluate the performance analytically and hence design efficient numerical solutions.


To make problem (P1) more tractable, we first take a deeper look at the ergodic SE in \eqref{SEFormula}. 
Without instantaneous per-RB CSI, the ergodic SE is statistically the same for different RB index $s$, under random channel fading.
Therefore, we simplify the per-RB allocation (based on $a_{s,u,q}$) into per-slot RB ratio optimization (based on $\rho_{u,q}$ which is the ratio of total RBs allocated to UE $u$ in slot $q$).
Second, consider practical power allocation and AIRS beamforming schemes. For example, consider equal power allocation among the RBs in each slot, i.e. $p_{s,q} = P_\text{BS}/S, \forall q \in \mathcal{Q}$.
Consider the low-complexity AIRS beamforming scheme based on MCCM as discussed in Section \ref{beamforming}.
Under these given setups, we can then evaluate the throughput $R_u$ and perform scheduling optimization.
Problem (P1) is thereby transformed into the following form:
\begin{align}
\mathrm{(P2)}:& \underset{
	\begin{subarray}{c}
	\rho_{u,q}, \mu_{i,u,q}, \\
	u \in \mathcal{U}, q \in \mathcal{Q}, i\in\mathcal{I}
	\end{subarray}
}{\max} \quad \min_{u \in \mathcal{U}} R_u \notag\\
\text{s.t.}\quad& 0 \leq \rho_{u,q} \leq 1, \forall u \in \mathcal{U}, q \in \mathcal{Q},  \label{SucarrierConstraintInP2} \\
& \textstyle\sum\nolimits_{u\in\mathcal{U}} \rho_{u,q} \leq 1, \forall q \in \mathcal{Q},  \label{SucarrierConstraint2InP2} \\
& \eqref{MatchConstraintInP1} \sim \eqref{MatchVarInP1} \notag.
\end{align}
Under practically large number of total RBs, the RB ratio $\rho_{u,q}$ can be approximated as a continuous variable, thus significantly reducing the complexity for discrete per-RB allocation.
However, the UE-to-AIRS association constraints \eqref{MatchConstraintInP1} $\sim$ \eqref{MatchVarInP1} are still discrete.
Moreover, the ergodic throughput $R_u$ needs effective ways to be evaluated and then improved.

\subsection{Ergodic SE Evaluation and CKM Solution}

We follow the CKM philosophy to accumulate historical channel/SE measurements into spatially correlated channel knowledge, and subsequently employ the CKM for ergodic SE prediction.
Specifically, assume that the UE positions could be obtained either through cellular or non-cellular assisted positioning methods\cite{CKMTutorial}. 
Moreover, similarly to \cite{PanFormula}, 
assume that for a certain UE $u$, the channel statistics (e.g., mean, variance, and/or empirical distribution function) for the link powers $| h_{\text{b},u,s} |^2 $, $| h_{\text{b},j,u,s} |^2$ and $\| \mathbf{h}_{i,u,s} \|^2$ could be obtained through statistical channel measurements.
In the meanwhile, under given UE-to-AIRS association, assume that the ergodic SE of UE $u$ served by AIRS $j_u$ could be estimated by statistical average of throughput measurements.


However, due to the large dimensions of system configurations (e.g., BS transmit power, AIRS amplification power, AIRS panel position/orientation, etc.),
it incurs large storage overhead to store the complete set of CKMs under various configuration conditions.
Moreover, it remains challenging to perform spatial interpolation/nonlinear inference among these conditional CKMs, in order to predict ergodic SEs in under-sampled locations/conditions.
These challenges motivate our proposed neural CKM in the next section.

\section{Neural CKM}\label{NeuralCKMSection}
In this section, the task of predicting the ergodic SE is decomposed into two sub-tasks, i.e., 1) predicting the per-link power statistics based on given UE positions and system configurations; and 2) composing the link power statistics to infer the ergodic SE under given UE-to-AIRS association.
To capture the highly nonlinear function mappings associated with these two sub-tasks, we design two cascaded deep neural networks (DNNs), with Transformer\cite{viT} as the core building blocks for flexible input encoding and output mapping.

\subsection{Overall Neural CKM Design}
\begin{figure} [t]
	\centering
	\includegraphics[width=0.85\linewidth,  trim=0.0 0.0 0.0 3.0, clip]{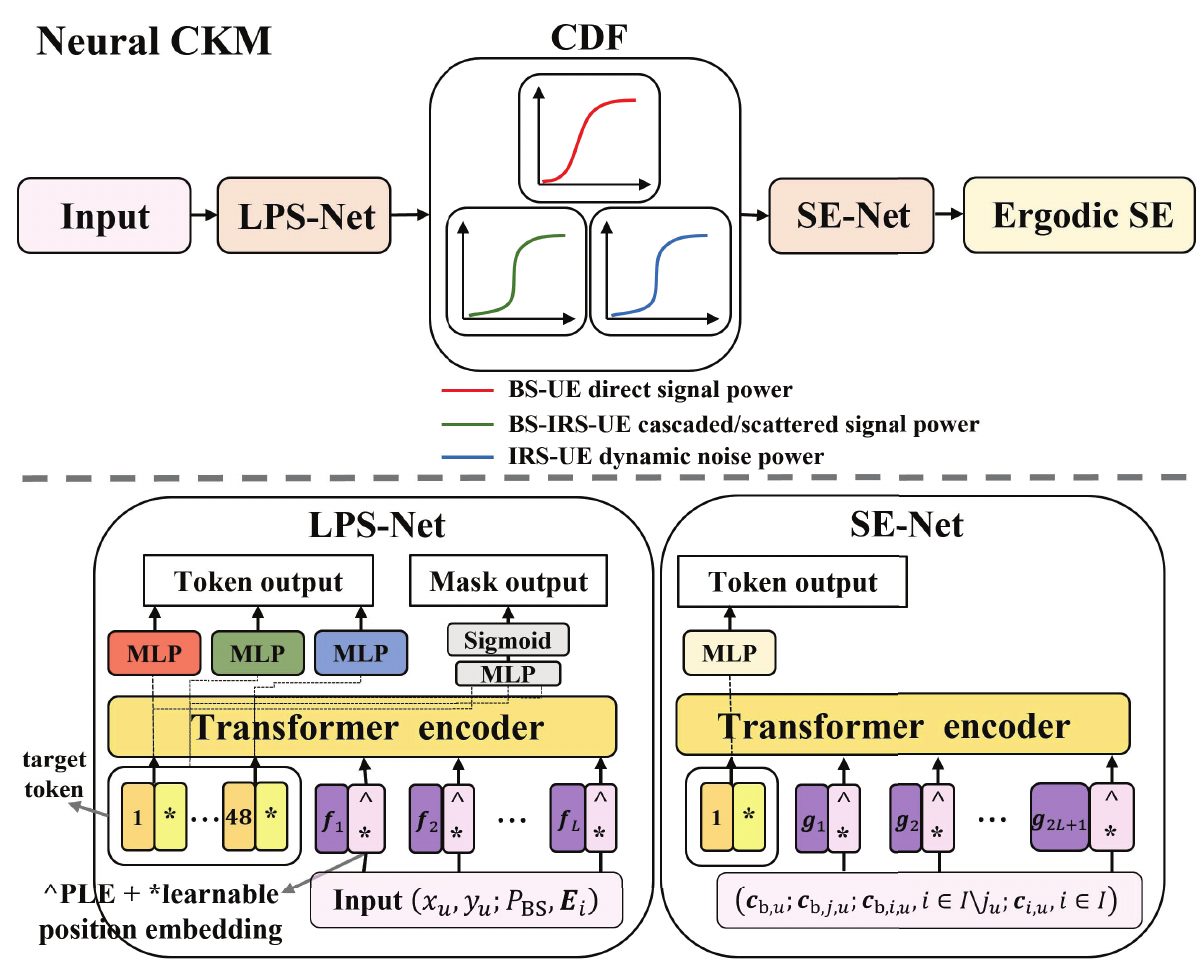}
	\caption{Neural CKM to predict link power statistics and ergodic SE.\vspace{-4ex}}
	\label{CKM}
\end{figure}

The overall Neural CKM design is illustrated in Fig. \ref{CKM}, which addresses the above two sub-tasks using two cascaded DNNs, i.e., the Link Power Statistics (LPS)-Net, and SE-Net.
These two DNNs share similar frameworks based on Transformer, though with different input/output module designs.

The LPS-Net is responsible for predicting the link power statistics, e.g., in the form of cumulative density function (CDF), for the direct BS-UE signal, BS-AIRS-UE cascaded/scattered signals, and IRS-UE dynamic noise for a given UE $u$ with known position and system configurations.
Specifically, each of the above link power CDF is represented by uniformly sampled 16 quantile points, and expressed in vector form as $\mathbf{c}_{\text{b},u}, \mathbf{c}_{\text{b},i,u}, \mathbf{c}_{i,u} \in \mathbb{R}^{16\times 1}$, respectively.
Let $\mathbf{P}_{u,i} \triangleq \left( \mathbf{c}_{\text{b},u}, \mathbf{c}_{\text{b},i,u}, \mathbf{c}_{i,u}\right) \in \mathbb{R}^{48\times 1}$ be the output of LPS-Net, which is a function of the UE position $(x_u,y_u)$ conditioned on the BS transmit power and the parameters $\mathbf{E}_{i}$ of AIRS $i$, i.e.,
\begin{equation}
\mathbf{P}_{u,i} \triangleq f_\textrm{LPS}\left(x_u, y_u;  P_\text{BS}, \mathbf{E}_i\right),
\end{equation}
where $\mathbf{E}_{i} \triangleq \left(x_i, y_i, H_i, G, \omega_{x,i}, \omega_{y,i}, \omega_{z,i}, W_\text{Y}, W_\text{Z}, P_\text{A}, \sigma_{\text{v}}^2, b_i\right)$, with $b_i$ = 1/0 indicating cascaded/scattered signal, respectively.

After the per-link power statistics are obtained by LPS-Net, they are then composed based on a given UE-to-AIRS association schedule to predict the ergodic SE.
Specifically, the input of SE-Net includes the direct BS-UE link power CDF $\mathbf{c}_{\text{b},u}$, the cascaded signal power CDF $\mathbf{c}_{\text{b},j_u,u}$ from the serving AIRS $j_u$, the scattered signal power CDF $\mathbf{c}_{\text{b},i,u}$ from other non-serving AIRSs, and the dynamic noise power CDF $\mathbf{c}_{i,u}$ from all AIRSs.
As a result, the ergodic SE is predicted by
\begin{equation}\label{SE-Net-formula}
\overline{\eta}_{u} = f_\textrm{SE} \left(\mathbf{c}_{\text{b},u}; \mathbf{c}_{\text{b},j_u,u}; \mathbf{c}_{\text{b},i,u}, i\in\mathcal{I}\setminus j_u; \mathbf{c}_{i,u},i\in\mathcal{I}\right).
\end{equation}

\subsection{Design of LPS-Net}

The framework of LPS-Net is shown in Fig. \ref{CKM}. 
Firstly, the input features of LPS-Net are scalar variables each with different meanings and ranges, which need proper encoding to map into the same dimensional feature space.
To this end, we employ the piecewise-linear encoding (PLE)\cite{PLE} to map each feature into a 256-dimensional (D) vector, which can then be treated as a token for subsequent Transformer blocks.
As a result, for a total of $L$ input features $x_l, l\in\mathcal{L}\triangleq\{ 1,\cdots,L\}$, we can obtain $L$ feature tokens $\mathbf{f}_l(x_l)\in \mathbb{R}^{1\times 256},l\in\mathcal{L}$.

On the other hand, in order to output the link power CDFs $\mathbf{P}_{u,i} \in \mathbb{R}^{48\times 1}$, we place 48 learnable target tokens $\mathbf{X}_\text{t}\in\mathbb{R}^{48\times 256}$ at the input.
Moreover, since attention-based models are inherently invariant to the order of input tokens, learnable position embeddings $\mathbf{E}_\text{pos}\in\mathbb{R}^{(48+L)\times 256}$ are added for distinction.
Then the overall input tokens are given by
\begin{equation}
    \mathbf{X} = \left[\mathbf{X}_\text{t}; \mathbf{f}_1(x_1), \cdots, \mathbf{f}_L(x_L)\right] + \mathbf{E}_\text{pos}.
\end{equation}

Secondly, the input tokens are processed by the Transformer encoder, comprising $A$ blocks of multi-head self-attention (MSA), MLP, layer normalization (LN), and residual connections\cite{viT}. 
Notably, the feature tokens  $\mathbf{f}_l(x_l),l\in\mathcal{L}$ attend to each other block-by-block to extract high-level feature correlations, while the learnable target tokens attend to the feature tokens along the blocks to extract key information related to link power CDFs.
Next, the target tokens processed by Transformer are fed to three MLPs to output the three CDFs  $\mathbf{c}_{\text{b},u}, \mathbf{c}_{\text{b},i,u}, \mathbf{c}_{i,u}$, respectively.
In addition, to deal with abnormal labels that may include negative infinity power values in dB,
the target tokens are also passed to a separate MLP followed by Sigmoid activations to output a 48-D probability mask to tell the validity of each CDF quantile point.

Finally, the overall loss function consists of three parts, i.e., 1) a smooth L1 loss function $f_{\delta_{\text{LPS}}}$ with parameter $\delta_{\text{LPS}}\in (0,1)$ for those valid quantile points, 2) a slope-MAE loss function $f_{\text{s-MAE}}$ to measure the average difference of function slopes between two CDFs, and 3) a binary cross-entropy (BCE) loss function $f_{\text{BCE}}$ for the masking module.
Therefore, we have
\begin{equation}
Loss_{\text{LPS-Net}} \triangleq f_{\delta_{\text{LPS}}} + \gamma \cdot f_{\text{s-MAE}} + \eta \cdot f_{\text{BCE}},
\end{equation}
where $\gamma$ and $\eta$ are constant weights.

\subsection{Design of SE-Net}

Unlike LPS-Net, SE-Net's inputs consist of $2I+1$ discrete CDFs in total, each of which is represented by 16 quantile points.
Firstly, we encode each CDF as a 256-D token, by encoding each quantile using 16-D PLE and concatenating them together.
We thus obtain $2I+1$ CDF tokens, each of which is further added by a position embedding on the corresponding category index, i.e., index 1 for direct link, index 2 for AIRS cascaded signal, index 3 for AIRS scattered signal, and index 4 for AIRS dynamic noise, as illustrated in Fig. \ref{CKM}.
In addition, a 256-D target token with learnable position embedding is also placed at the input for ergodic SE prediction.

Secondly, the overall $2I+2$ input tokens are processed by the Transformer blocks to extract high-level correlations, after which the target token is fed to a MLP layer to predict the ergodic SE. Smooth L1 loss with $\delta_{\text{SE}}\in(0,1)$ is adopted, i.e.,
\begin{equation}
Loss_{\text{SE-Net}} \triangleq f_{\delta_{\text{LPS}}}(\bar{\eta},\hat{\bar{\eta}}),
\end{equation}
which compares the labeled/predicted ergodic SEs $\bar{\eta}$ and $\hat{\bar{\eta}}$.

\section{Neural CKM based Scheduling Optimization} 

Based on our proposed neural CKM, the ergodic SE of UE $u$ under given UE-to-AIRS association can be predicted by SE-Net, and hence its throughput $R_u$ can be obtained by \eqref{Ru} under given schedule.
This facilitates efficient numerical evaluation of the objective in Problem (P2), thus paving ways for our further design of scheduling optimization.

Specifically, for our considered max-min ergodic throughput objective in Problem (P2), the available resources (i.e., AIRSs and RBs/slots) need to be properly allocated to strike a balance among the UEs, considering their different achievable SE and the AIRS time sharing constraints.
To this end, we propose a three-stage algorithm named \textit{Stable Matching-Iterative Balancing (SM-IB)}. Stage 1 initializes the group of UEs allocated in each slot based on stable matching between UE clusters and the AIRSs. 
Stage 2 maximizes the local minimum throughput in each slot by iteratively tuning the RB allocation ratios among the UEs. 
Stage 3 maximizes the global minimum throughput across the slots by iteratively swapping UEs between slots with currently minimum/maximum per-UE throughput.
Stage 2 is nested within Stage 3, both of which are performed iteratively to balance the throughput across all UEs. 


%
%

\subsection{Initial UE Grouping Based on Stable Matching}
The mission in Stage 1 is to coarsely balance the ergodic SE of UEs grouped into different slots.
The basic idea is to first cluster the UEs based on their preference on the AIRSs, and then group the UEs with poor/good SE into one slot to balance their RB requirements.

Specifically, consider the non-trivial case where the number of UEs $U$ is larger than the number of AIRSs $I$. 
Based on SE-Net, each UE $u$ first inquires about their expected ergodic SE when served by each of the AIRSs, which comprises an SE vector $\bar{\boldsymbol{\eta}}_{u}\in\mathbb{R}^{I\times 1}$.
Based on the SE vector $\bar{\boldsymbol{\eta}}_{u}$, $u\in\mathcal{U}$, the constrained K-Means (cKMeans) algorithm
is applied to classify all UEs into $I$ clusters, each with at least $U_0$ UEs whereby $\lfloor U/I\rfloor\leq U_0\leq Q$. 
These $I$ clusters are then one-to-one matched to the $I$ AIRSs based on mutual SE preference, using the Gale-Shapley (GS) algorithm \cite{GSalogrithm} to form stable matching.
Next, within each cluster, UEs are sorted by ascending order of their expected SE when served by the BS only.
Then the top $Q$ bottleneck UEs in each cluster are evenly assigned to the $Q$ slots and served by the corresponding AIRS.
All the remaining UEs are served by the BS only, and distributed evenly across the slots based on descending order of their expected SE.

\subsection{Max-Min Throughput in Each Slot}
Note that the initial UE grouping above aims for coarse-grained UE-to-AIRS/slot mapping.
Next, in Stage 2, fine-grained scheduling optimization for max-min throughput within each slot is performed, including refined UE-to-AIRS matching and RB ratio allocation.
Due to the high complexity of joint optimization, we leverage alternating optimization (AO) to decompose Problem (P2) into two subproblems.

Subproblem 1 optimizes the RB ratio allocation under given UE-to-AIRS matching in each slot.
Recall that UE $u$'s ergodic throughput in \eqref{Ru} is a monotonic function of its allocated RB ratio $\rho_{u,q}$.
For max-min throughput in each slot, we propose an Iterative Balancing (IB) algorithm that iteratively picks up two UEs with the best/worst throughput, and adjusts their RB ratios to reduce their throughput gap. The process continues until the throughput gap is smaller than a threshold $\epsilon$, which is of complexity $\mathcal{O}\left(\log_{2}( 1/\epsilon)\right)$.

In turn, given the RB allocation ratios, subproblem 2 refines the UE-to-AIRS matching in a given slot, which is solved using the GS algorithm to form stable one-to-one matching, based on the expected ergodic SE of each UE when served by a given AIRS or by the BS only.
Since the number of AIRSs is typically smaller than the number of UEs in a given slot, 
AIRSs act as the proposing side while UEs act as the accepting side, according to \cite{GSalogrithm}.
Since each AIRS can make at most one proposal to each UE, the complexity of the GS algorithm is $\mathcal{O}\left(IU\right)$.
After a stable matching is found, unmatched UEs are served by the BS only.

The above two subproblems are alternatingly solved for $N_{\max}$ iterations until local max-min throughput in a given slot is found. Since both the UE-to-AIRS matching process and the RB ratio tuning process monotonically improve the minimum throughput, convergence is guaranteed, with the overall complexity given by $\mathcal{O}\left(N_{\max}\big(IU+\log_{2}\left( 1/\epsilon \right)\big)\right)$.

\subsection{Max-Min Throughput across Slots}
Stage 2 pursues per-slot max-min throughput under given UE grouping in each slot, which leaves room for improvement if cross-slot UE grouping could be refined.
To this end, in Stage 3, we extend the IB algorithm to balance the minimum throughput across different slots by iteratively swapping UEs.

Specifically, in each iteration, the two slots corresponding to the minimum and maximum common throughput are selected, denoted as $q_{\min} $and $q_{\max}$, respectively.
Then, one UE from each of these two slots is chosen to swap.
To improve the minimum throughput, the UE in $q_{\min}$ with the largest occupied RB ratio (and hence the lowest SE) is exchanged with a UE from $q_{\max}$.
After swapping, the RB allocation ratios of the remaining UEs in slot $q_{\min}$ or $q_{\max}$ are first scaled proportionally to fit the total bandwidth.
Then, the max-min process in Stage 2 is performed in these two slots.
Note that a UE swap is actually adopted only if the minimum throughput could be improved in the above process, which guarantees monotonic convergence.
As such, Stage 3 is performed iteratively until the common throughput gap across different slots is smaller than a threshold $\xi$, which requires $\mathcal{O}\left(\log_{2}\left( 1/\xi \right)\right)$ iterations.
As a result, the proposed SM-IB algorithm iteratively solves for the max-min throughput problem in (P2), with the overall complexity of $\mathcal{O}\left(\log_{2}\left( 1/\xi \right)\cdot N_{\max}\big(IU+\log_{2}\left( 1/\epsilon \right)\big)\right)$.

\section{Numerical Results}
Numerical results are provided for performance evaluation.
The following parameters are used if not mentioned otherwise: 
$f_\text{c}=3.5$ GHz, $B = 20$ MHz, $P_\text{BS} = 10$ dBm, $P_\text{A} = 10$ dBm, $G_\text{i} = 6$ dBi, $N_\text{v} = -160$ dBm/Hz and $N_\text{0} = -174$ dBm/Hz.

\subsection{Performance of Different Phase Optimization Schemes} 
Our neural CKM method can be applied to any given AIRS phase optimization schemes.
Here we justify the adoption of MCCM-based method by comparing with three benchmark schemes, including 1) the capacity characterization method in \cite{CapacityZhangShuoWen} which serves as an upper bound, 2) LoS beamforming that steers the AIRS beam directly towards the UE position, and 3) random phase.
The ergodic SE and running time are evaluated.
Monte Carlo (MC) simulation is conducted in a single BS-AIRS-UE scenario,
with 50 realizations of large-scale parameters, each with 500 instances of small-scale fading. 



\begin{figure} [t]
\centering
\includegraphics[width=0.65\linewidth, trim=0.0 0.0 0.0 0.0, clip]{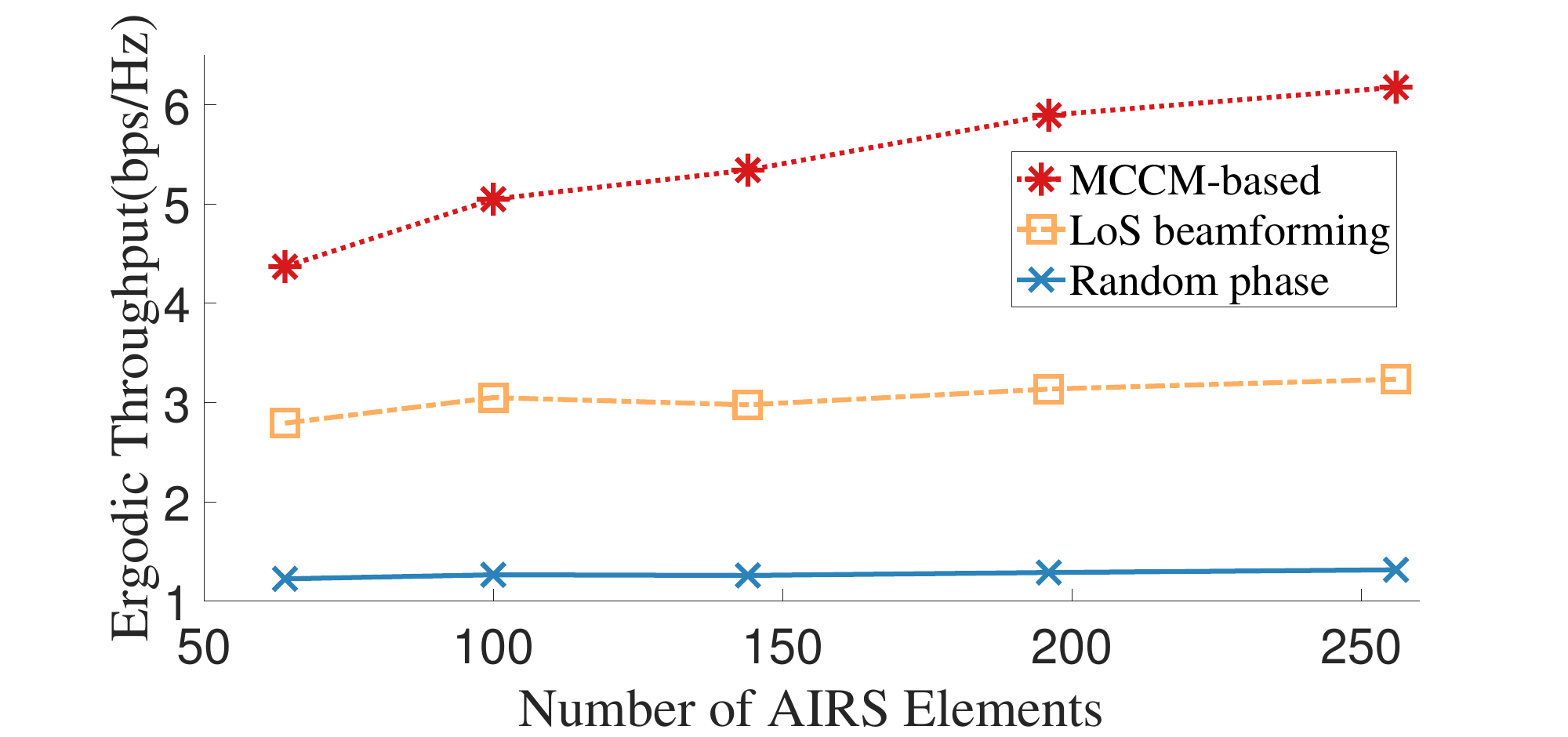}
\caption{Ergodic throughput under different number of AIRS elements.\vspace{-2ex}}
\label{Phase_comp}
\end{figure}

First, the performance comparison with \cite{CapacityZhangShuoWen} is conducted for an AIRS with $4 \times 4$ elements. 
It is found that the MCCM-based method achieves an ergodic SE close to the upper bound achieved by \cite{CapacityZhangShuoWen} (3.233 versus 3.357 bps/Hz), yet with significantly reduced time (0.01 versus 351 seconds).
Hence, the MCCM-based scheme achieves a better trade-off between SE performance and practical implementation.
On the other hand, for the low-complexity LoS beamforming scheme and the random phase baseline, the ergodic SE with larger number of AIRS elements is shown in Fig. \ref{Phase_comp}.
It can be seen that the MCCM-based method significantly outperforms the other two schemes. In particular, the MCCM captures richer correlations in multi-path/RBs/AIRS elements, and hence is better than LoS beamforming based on the UE location only.



\subsection{Effectiveness of Neural CKM}
To train the neural CKM, a dataset consisting of 10,000 samples is constructed based on MC simulations in QuaDRiGa.
The dataset is split into training, validation, and test sets in a ratio of 8:1:1. 
To mitigate potential random errors in the validation set, K-fold cross-validation is employed.
Besides, the encoder module adopts a 4-layer, 4-head attention architecture. The MLP module within the encoder consists of two linear layers, two dropout layers, and GELU activation functions.
The loss function parameters are given by $\delta_{\text{LPS}} = 0.5$, $\gamma = 0.2$, $\eta = 20.0$ and $\delta_{\text{SE}} = 1.0$.
The optimizer used is AdamW, with an initial learning rate of $1\text{e}^{-5}$ and weight decay of $1\text{e}^{-5}$. 
The neural CKM is implemented in PyTorch 2.3.1 and trained on a single Nvidia GeForce RTX 4090 GPU with 24GB of memory.


To validate the effectiveness of our Transformer-based neural CKM design, we consider pure MLP based design for LPS-Net and LSTM-based design for SE-Net\footnote{LSTM is considered for SE-Net to handle varying number of AIRSs.} as baselines.  
The hyperparameters of both baselines are tuned with best effort for prediction accuracy.
The performance comparison is summarized in Table \ref{tab:NeuralCKM}, including NN parameter count, average inference time per sample, MAE and mean relative error (MRE) for SE prediction, and quantile MAE (QMAE) and quantile MRE (QMRE) for CDF prediction.
It can be seen that the proposed neural CKM design achieves higher prediction accuracy compared with baselines, with similar inference time.
Moreover, the per-sample inference time is in milliseconds and thus facilitates efficient SE prediction and further optimization.

\begin{table}[htbp]
\caption{Performance comparison of Neural CKM and benchmarks.}
\centering
\resizebox{\linewidth}{!}{
\begin{tabular}{lccccccc}
\toprule
\textbf{Scheme} & \text{Network} & $N_\text{para.}$ & $\bar{T} (s)$ & \text{QMAE} (dB)& \text{QMRE} ($\%$)& \text{MAE} (bps/Hz)& \text{MRE} ($\%$)\\
\midrule
Proposed & LPS-Net & 3648004 & 0.0026 & 1.8734 & 0.92  & / & / \\
Baseline & MLP & 4811390 & 0.0012 & 7.1802 & 2.94  & / & / \\
Proposed & SE-Net & 3237377 & 0.0041 & / & / & 0.2093 & 8.65 \\
Baseline & LSTM & 603905 & 0.0040 & / & / & 0.2220 & 11.08 \\
\bottomrule
\end{tabular}}
\label{tab:NeuralCKM}
\end{table}

\subsection{Effectiveness of SM-IB Algorithm for Max-Min Scheduling}

Based on the trained neural CKM, we can further evaluate the performance of the proposed SM-IB algorithm for max-min scheduling.
A multi-user system assisted by multiple (e.g., 2 or 6) AIRSs with $W = 12 \times 12$ elements is considered as an example, as illustrated in Fig. \ref{Fig:system_model}.
Two benchmark schemes are considered, i.e., 1) the performance upper bound obtained via the Gurobi solver for solving (P2), and 2) the performance lower bound with randomized UE-to-AIRS/slot scheduling and equal RB ratio allocation per slot.

\begin{figure} [t]
\centering
\includegraphics[width=0.65\linewidth, trim=0.0 0.0 0.0 0.0, clip]{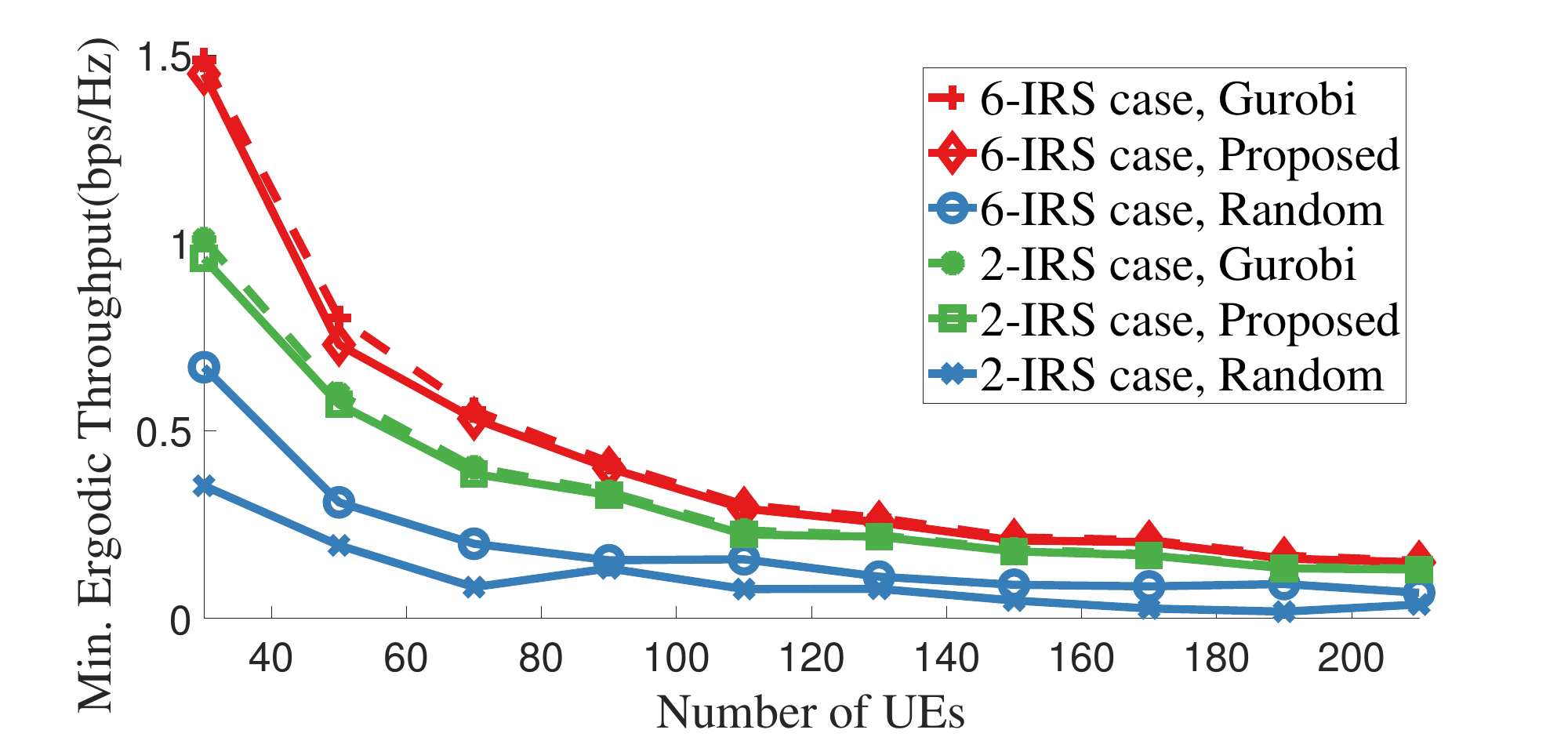}
\caption{Minimum ergodic throughput per UE under different schemes.\vspace{-2ex}}
\label{Fig:Algorithm_comp}
\end{figure}

The achieved minimum ergodic throughput per UE is shown in Fig. \ref{Fig:Algorithm_comp}, under different number of AIRSs and UEs.
It can be seen that the proposed algorithm achieves a max-min ergodic throughput close to that of the upper bound, and significantly outperforms the random scheduling baseline.
Moreover, as shown in Table \ref{tab:time_comparison}, the average running time for Gurobi grows significantly as the number of UEs increases due to the inherent combinatorial complexity, while our propose SM-IB algorithm maintains a low level of complexity and running time.


\begin{table}[htbp]
\caption{Average running time (s) for solving (P2) using Gurobi or the proposed scheme for the 6-IRS case.}
\label{tab:time_comparison}
\centering
\resizebox{\linewidth}{!}{
\begin{tabular}{l|ccccccccccc}
\toprule
\diagbox{Scheme}{$U$} & 30 & 50 & 70 & 90 & 110 & 130 & 150 & 170 & 190 & 210\\
\midrule
Gurobi & 13.4 & 28.0 & 61.4 & 109.1 & 141.2 & 256.1 & 194.8 & 325.6 & 1080.6 & 1497.2 \\
Proposed & 0.059 & 0.085 & 0.089 & 0.170 & 0.134 & 0.090 & 0.126 & 0.182 & 0.186 & 0.344 \\
\bottomrule
\end{tabular}}
\end{table}

\section{Conclusions}
This paper addressed the critical scheduling challenge in multi-user wireless networks aided by active IRSs, specifically targeting the complexity arising from real-time channel information acquisition and resource optimization. 
A novel neural CKM method, featuring dedicated Transformer-based designs of LPS-Net and SE-Net, is proposed to predict ergodic SE based on historical channel/throughput measurements and spatial correlations.
Furthermore, a low-complexity SM-IB algorithm is proposed for efficient scheduling. 
Numerical results verify that the proposed neural CKM significantly enhances prediction accuracy and computational efficiency, while the SM-IB algorithm effectively achieves near-optimal max-min throughput with greatly reduced complexity.

\bibliography{IEEEabrv,bibliography}

\begin{thebibliography}{10}
\providecommand{\url}[1]{#1}
\csname url@samestyle\endcsname
\providecommand{\newblock}{\relax}
\providecommand{\bibinfo}[2]{#2}
\providecommand{\BIBentrySTDinterwordspacing}{\spaceskip=0pt\relax}
\providecommand{\BIBentryALTinterwordstretchfactor}{4}
\providecommand{\BIBentryALTinterwordspacing}{\spaceskip=\fontdimen2\font plus
\BIBentryALTinterwordstretchfactor\fontdimen3\font minus
  \fontdimen4\font\relax}
\providecommand{\BIBforeignlanguage}[2]{{%
\expandafter\ifx\csname l@#1\endcsname\relax
\typeout{** WARNING: IEEEtran.bst: No hyphenation pattern has been}%
\typeout{** loaded for the language `#1'. Using the pattern for}%
\typeout{** the default language instead.}%
\else
\language=\csname l@#1\endcsname
\fi
#2}}
\providecommand{\BIBdecl}{\relax}
\BIBdecl

\bibitem{RecentAdvances}
Q.~Wu, B.~Zheng, C.~You \emph{et~al.}, ``Intelligent surfaces empowered
  wireless network: {Recent} advances and the road to {6G},'' \emph{Proc.
  IEEE}, vol. 112, no.~7, pp. 724--763, 2024.

\bibitem{ZhengBeiXiongNOMAorOMA}
B.~Zheng, Q.~Wu, and R.~Zhang, ``Intelligent reflecting surface-assisted
  multiple access with user pairing: {NOMA} or {OMA}?'' \emph{IEEE Commun.
  Lett.}, vol.~24, no.~4, pp. 753--757, Apr. 2020.

\bibitem{zhang2021active}
Z.~Zhang, L.~Dai, X.~Chen, C.~Liu, F.~Yang, R.~Schober, and H.~V. Poor,
  ``Active {RIS} vs. passive {RIS}: Which will prevail in {6G}?'' \emph{IEEE
  Trans. Commun.}, vol.~71, no.~3, pp. 1707--1725, 2022.

\bibitem{ActiveIRSOpenIssue}
Z.~Kang, C.~You, and R.~Zhang, ``Active-{IRS}-aided wireless communication:
  {Fundamentals}, designs and open issues,'' \emph{IEEE Wireless Commun.},
  vol.~31, no.~3, pp. 368--374, 2024.

\bibitem{PanFormula}
Z.~Peng, X.~Liu, C.~Pan \emph{et~al.}, ``Multi-pair {D2D} communications aided
  by an active {RIS} over spatially correlated channels with phase noise,''
  \emph{IEEE Wireless Commun. Lett.}, vol.~11, no.~10, pp. 2090--2094, 2022.

\bibitem{Dongsheng}
D.~Fu \emph{et~al.}, ``Site-specific deployment optimization of intelligent
  reflecting surface for coverage enhancement,'' in \emph{VTC-Spring}, 2024,
  pp. 01--07.

\bibitem{OFDMA-schedule-4}
Y.~Yang, S.~Zhang, and R.~Zhang, ``{IRS}-enhanced {OFDMA}: Joint resource
  allocation and passive beamforming optimization,'' \emph{IEEE Wireless
  Commun. Lett.}, vol.~9, no.~6, pp. 760--764, 2020.

\bibitem{OFDMA-schedule-5}
W.~Wu, F.~Yang, F.~Zhou, Q.~Wu, and R.~Q. Hu, ``Intelligent resource allocation
  for {IRS}-enhanced {OFDM} communication systems: {A} hybrid deep
  reinforcement learning approach,'' \emph{IEEE Trans. Wireless Commun.},
  vol.~22, no.~6, pp. 4028--4042, 2023.

\bibitem{NOMA-schedule-1}
X.~Mu, Y.~Liu, L.~Guo, J.~Lin, and N.~Al-Dhahir, ``Capacity and optimal
  resource allocation for {IRS}-assisted multi-user communication systems,''
  \emph{IEEE Trans. Commun.}, vol.~69, no.~6, pp. 3771--3786, 2021.

\bibitem{CKMTutorial}
Y.~Zeng, J.~Chen, J.~Xu \emph{et~al.}, ``A tutorial on environment-aware
  communications via channel knowledge map for 6g,'' \emph{IEEE Commun. Surveys
  Tuts.}, vol.~26, no.~3, pp. 1478--1519, 2024.

\bibitem{CKM_IRSbeamSelection}
D.~Ding, D.~Wu, Y.~Zeng, S.~Jin, and R.~Zhang, ``Environment-aware beam
  selection for {IRS}-aided communication with channel knowledge map,'' in
  \emph{IEEE GLOBECOM Wkshps.}, 2021, pp. 1--6.

\bibitem{CKM_ActivePassive}
E.~Moeen~Taghavi \emph{et~al.}, ``Environment-aware joint active/passive
  beamforming for {RIS}-aided communications leveraging channel knowledge
  map,'' \emph{IEEE Commun. Lett.}, vol.~27, no.~7, pp. 1824--1828, 2023.

\bibitem{viT}
A.~Dosovitskiy \emph{et~al.}, ``An image is worth 16x16 words: {Transformers}
  for image recognition at scale,'' in \emph{Proc. ICLR}, 2021.

\bibitem{TransShiJinPathlossmodeling}
W.~Tang \emph{et~al.}, ``Wireless communications with reconfigurable
  intelligent surface: Path loss modeling and experimental measurement,''
  \emph{IEEE Trans. Wireless Commun.}, vol.~20, no.~1, pp. 421--439, 2021.

\bibitem{CapacityZhangShuoWen}
S.~Zhang and R.~Zhang, ``Capacity characterization for intelligent reflecting
  surface aided {MIMO} communication,'' \emph{IEEE J. Sel. Areas Commun.},
  vol.~38, no.~8, pp. 1823--1838, 2020.

\bibitem{JiangTaoMCCM}
Y.~Chen, D.~Chen, and T.~Jiang, ``Beam-squint mitigating in reconfigurable
  intelligent surface aided wideband mmwave communications,'' in \emph{Proc.
  IEEE Wireless Commun. and Net. Conf. (WCNC)}, 2021, pp. 1--6.

\bibitem{PLE}
Y.~Gorishniy \emph{et~al.}, ``On embeddings for numerical features in tabular
  deep learning,'' in \emph{Proc. NIPS}, 2022.

\bibitem{GSalogrithm}
S.~Bayat \emph{et~al.}, ``Matching theory: Applications in wireless
  communications,'' \emph{IEEE Sig. Proc. Mag.}, vol.~33, no.~6, pp. 103--122,
  2016.

\end{thebibliography}

\end{document}